\documentclass[a4paper]{jpconf}
\usepackage{graphicx}
\usepackage{lineno}

\begin{document}
\title{Summary of recent experimental results on strangeness production}


\author{Alexander Kalweit}

\address{European Organization for Nuclear Research (CERN), Geneva, Switzerland}

\ead{Alexander.Philipp.Kalweit@cern.ch}

\begin{abstract}
This article summarises the highlights of the recent experimental findings on strangeness production presented at the 16th edition of the {\it International Conference on Strangeness in Quark Matter} in Berkeley. Results obtained by eight large experimental collaborations (ALICE, ATLAS, CMS, HADES, LHCb, NA-61, PHENIX, STAR) spanning a large range in centre-of-mass energy and a variety of collision systems were presented at the conference. The article does not aim at being a complete review, but rather at connecting the experimental highlights of the different collaborations and at pointing towards questions which should be addressed by these experiments in future.

\end{abstract}

\section{Introduction}

Historically, the interest in strangeness production in heavy-ion collisions dates back to the original proposal of strangeness enhancement as a signature of QGP formation~\cite{Rafelski:1982pu}. In a more modern view, the formation of strange particles in relativistic nucleus-nucleus (AA) collisions is described together with all other light flavour hadrons (hadrons consisting of $u$, $d$, and $s$ valence quarks) in the framework of statistical thermal models and hydrodynamics (see e.g.~\cite{BraunMunzinger:2003zd,Voloshin:2008dg}). 

\section{Nucleus-nucleus collisions at LHC and RHIC}

Light flavour hadrons of low transverse momentum define the collective behaviour of the fireball created in the collision of ultra-relativistic heavy ions. Even at the highest LHC energies about 98\% of all particles exhibit transverse momenta $p_{\rm T}< 2$~GeV/$c$~\cite{BelliniProceedings}. They are thus produced in a regime which is not easily accessible to perturbation theory approaches in QCD. However, the assumption that the medium created in the collision is in local thermodynamic equilibrium allows to describe the main features of bulk particle production with two basic concepts. Firstly, the $p_{\rm T}$-integrated particle yields follow the expectations from statistical-thermal models (chemical equilibrium). Secondly, the spectral shapes and azimuthal anisotropies can be explained by a common hydrodynamic expansion (kinetic equilibrium). One of the main goals in the current investigation of light flavour hadron production is to determine up to which precision this baseline model holds and to carefully investigate the deviations from it for the search of new phenomena.

\begin{figure}
\begin{center}
	\includegraphics[width=0.92\linewidth]{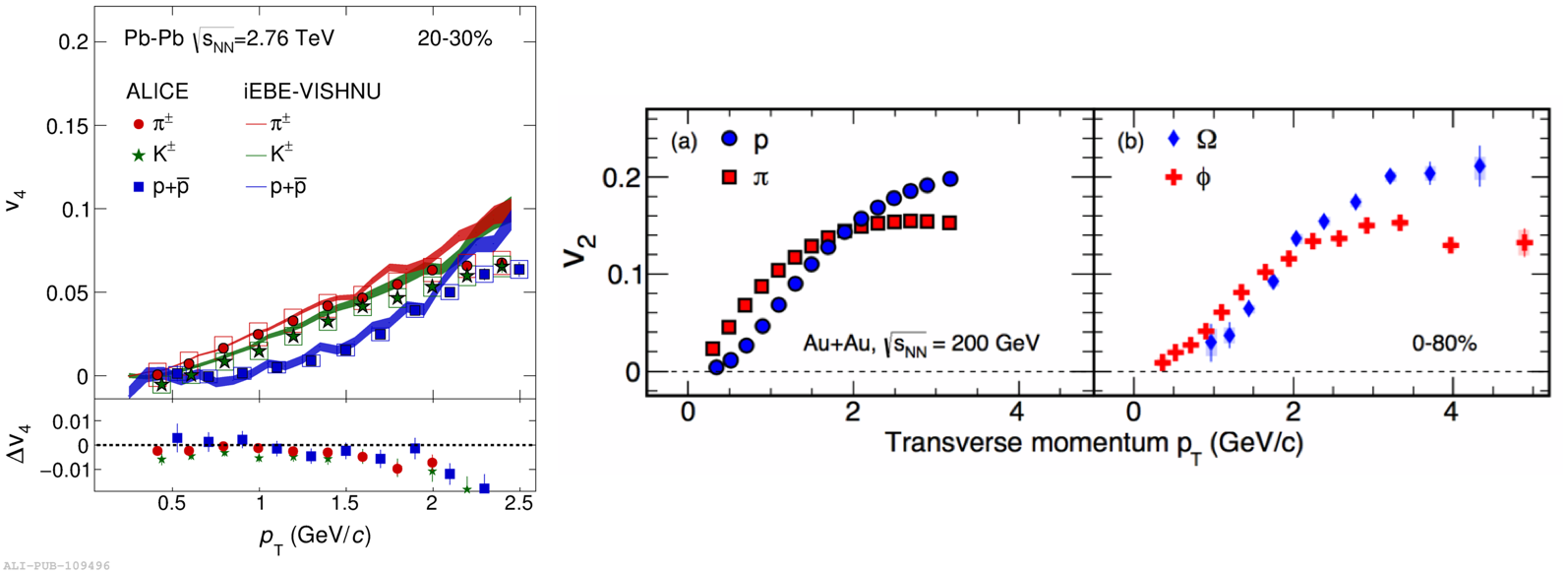}
\end{center}
\caption{\label{fig.:FlowMassOrdering} The $p_{\rm T}$-differential $v_4$ for $\pi$, K and p measured in 20-30\% Pb-Pb collisions at 2.76 TeV compared to expectations from hydrodynamics (left). The $v_2$ as function of $p_{\rm T}$ for $\pi$, p (panel a) and $\phi$, $\Omega$ (panel b) from minimum bias Au+Au collisions at 200~GeV for 0-80\% centrality (right). Figures taken from~\cite{Adam:2016nfo} (left) and from~\cite{Adamczyk:2015ukd} (right).
}
\end{figure}

The validity of the hydrodynamic picture has recently been confirmed at the highest available centre-of-mass energy by the ALICE collaboration with a measurement of the $v_2$, $v_3$, and $v_4$ flow harmonics for charged particles in Pb-Pb collisions at $\sqrt{s_{\rm NN}}=5.02$~TeV~\cite{Adam:2016izf}. One of the most distinct predictions of hydrodynamics is a mass ordering in the $p_{\rm T}$-dependent flow coefficients resulting from the interplay of radial and anisotropic flow: the common expansion velocity $\beta$ leads to stronger blue-shift for heavier particles following $p = m \cdot \beta\gamma$ where $m$ corresponds to the particle mass and $\gamma$ to the Lorentz factor. Figure~\ref{fig.:FlowMassOrdering}~(left) shows a striking confirmation of this prediction for the fourth order flow harmonic which was recently measured by ALICE~\cite{Adam:2016nfo}. The STAR Collaboration confirmed that the mass ordering for elliptic flow even holds for the triple-strange $\Omega$-Baryon as shown in Fig.~\ref{fig.:FlowMassOrdering}~(right). As the hadronic cross-sections for $\Omega$ and $\phi$ are significantly smaller than for the non-strange particles p and $\pi$, the results also indicate that a major part of the collectivity is already built up at the partonic stage.

\begin{figure}
\begin{center}
	\includegraphics[width=0.92\linewidth]{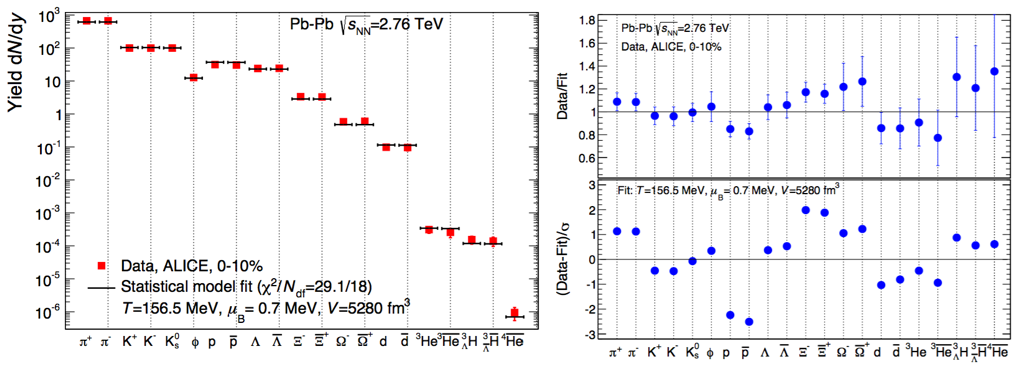}
\end{center}
\caption{\label{fig.:ThermalModel} Comparison of grand-canonical thermal model predictions with ALICE data for central Pb--Pb collisions. Figure taken from~\cite{AndronicProceedings}.
}
\end{figure}

The success of thermal-statistical models in describing particle yields over a wide range of energies in central heavy-ion collisions is also confirmed at LHC energies. The latest comparison of the complete set of particles available at LHC energies with the model prediction is shown in Fig.~\ref{fig.:ThermalModel}. The  yields of light flavour hadrons are described over nine orders of magnitude within 20\% with a common chemical freeze-out temperature of $T_{ch}~\approx$~156~MeV. Most notably, also the yields of light (anti-)(hyper-)nuclei show agreement with equilibrium thermal model predictions. The largest deviations are observed for protons ($\approx 2.8\sigma$) which would prefer a lower $T_{ch}$ and for $\Xi$ ($\approx 2.0\sigma$) which would prefer a higher $T_{ch}$. The question must be answered if these deviations are due to physical processes which are not taken into account in the current versions of thermal-statistical models or if the deviations are not significant and can be neglected by invoking the Occam's razor principle, giving preference to a model as simple as possible. One open issue was related to the p/$\pi$-ratio at top RHIC energies measured by STAR, that was found significantly higher than the ALICE measurement and whose value was thus indicating chemical freeze-out temperatures of approximately 163 MeV. This topic is now re-addressed with new data on proton production in which feed-down corrections are improved with the help of the STAR heavy-flavour tracker. The first results presented at the conference indeed show a lower p/$\pi$-ratio for prompt protons which is more consistent with the ALICE and PHENIX results~\cite{MizunoProceedings}. A careful future investigation of this topic is needed, because hadron resonance gas calculations on fluctuation observables start to deviate significantly from Lattice QCD for temperatures above 160 MeV~\cite{KarschProceedings}.

\begin{figure}[h]
\begin{minipage}{0.47\linewidth}
\includegraphics[width=\linewidth]{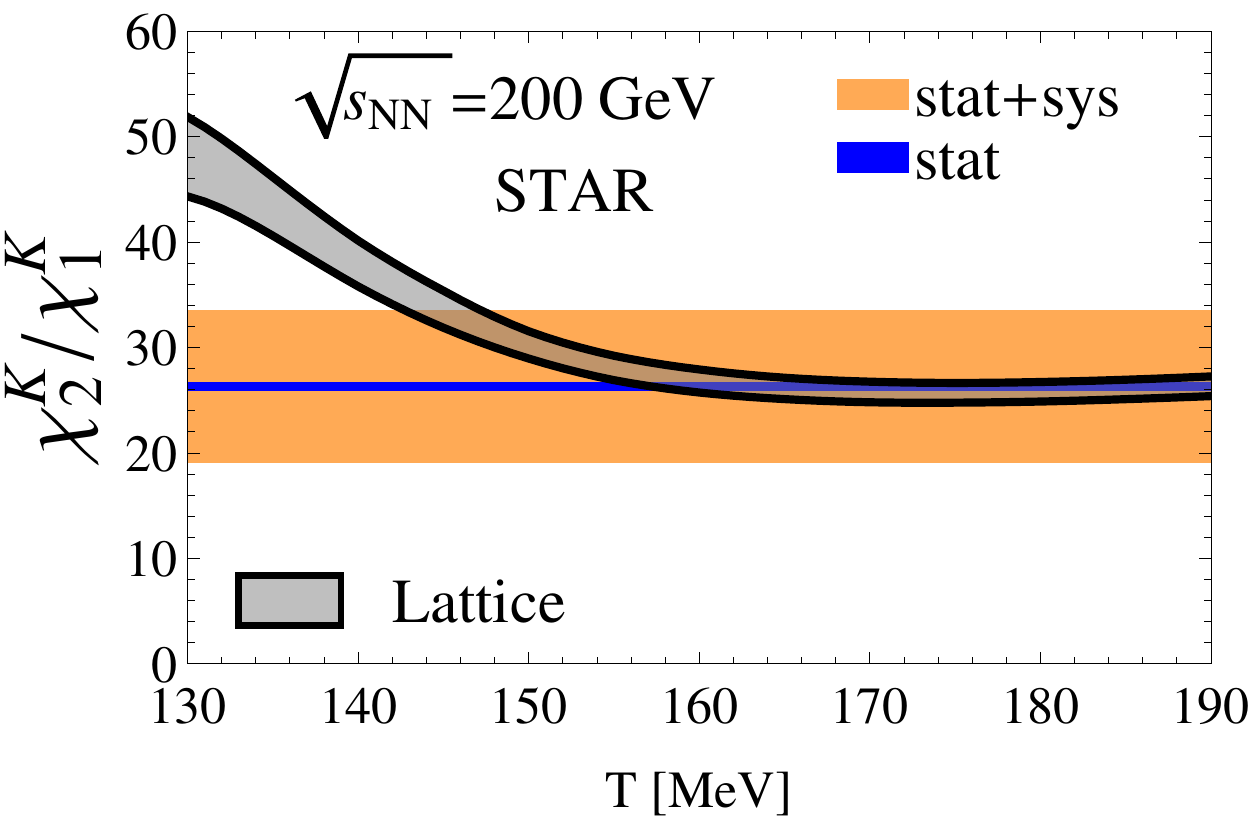}
\caption{\label{fig.:kaonFluct} Comparison of the net-kaon fluctuation measurement of the STAR collaboration with Lattice QCD predictions. Figure taken from~\cite{RattiProceedings,Noronha-Hostler:2016rpd}.}
\end{minipage}\hspace{2pc}%
\begin{minipage}{0.45\linewidth}
\includegraphics[width=\linewidth]{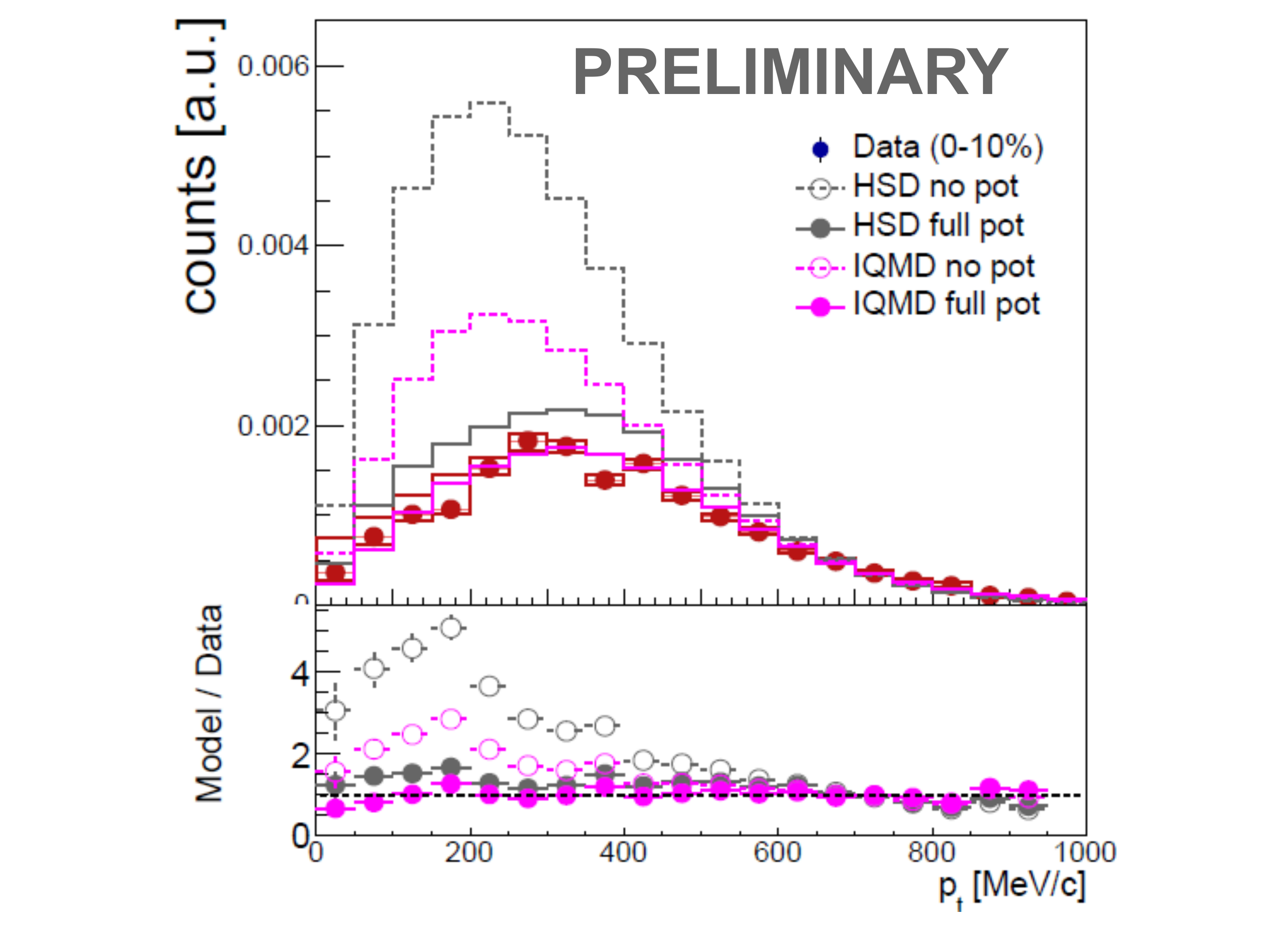}
\caption{\label{fig.:HadesTransport} Comparison of HADES kaon spectra  
with transport code calculations. 
Figure taken from~\cite{ScheibProceedings}.}
\end{minipage} 
\end{figure}

On the theoretical side, two main ideas are on the market as an explanation for the different p and $\Xi$ yield with respect to the thermal model at LHC energies: firstly, the concept that the chemical freeze-out temperatures for $s$ quarks could be higher than for $u$ and $d$ quarks and secondly the presence of inelastic collisions in the hadronic phase~\cite{Becattini:2016xct}. A potential difference in the chemical freeze-out temperature was originally motivated by Lattice QCD results from the Wuppertal-Budapest collaboration~\cite{Bellwied:2013cta}. In order to substantiate this claim, freeze-out parameters have previously been extracted from measurements of net-charge and net-baryon number fluctuations~\cite{Bazavov:2012vg,Borsanyi:2014ewa} and now first attempts have been made for net-strangeness fluctuations, but as shown in Fig.~\ref{fig.:kaonFluct}, the current experimental precision on net-kaon fluctuations is not sufficient to clarify if the freeze-out temperature for strange particles is higher than for non-strange particles. On the other hand, the dynamical picture of a freeze-out of a fraction of the partons in the system is not yet clear and also a clear ordering in the deviation of the particle yields from the thermal model is not observed: in principle, the triple strange $\Omega$-Baryon should show a larger deviation from a {\it common} chemical freeze-out temperature, but the data does not show the expected ordering ($\Omega > \Xi {\rm \; , \;} \phi >  \Lambda {\;\rm ,\; K}$) of the deviations. New data with increased precision for multi-strange particles from the LHC run 2 and the upgraded detectors at RHIC is needed in order to investigate the exact magnitude of the deviations and their ordering.

The effects of inelastic collisions in the hadronic phase are studied with UrQMD afterburners added to the thermal-statistical fits~\cite{Becattini:2016xct}. A significant improvement of the $\chi^2/n_{dof}$ is observed in this case together with an increase in $T_{ch}$. The increase in $T_{ch}$ leads at LHC energies to regimes of chemical freeze-out temperatures in which the hadron resonance gas already deviates significantly from Lattice QCD~\cite{KarschProceedings}. Future experimental work is needed to investigate precisely the centrality dependence of particle ratios such as p$/\pi$ and $\Xi/\pi$. While the plain statistical-thermal model predicts that particle ratios are saturated above approximately 60\% centrality, the lifetime of the hadronic phase is expected to increase with increasing centrality and thus also its influence on the particle ratios~\cite{Becattini:2014hla}. While the dependence of particle ratios on centrality in Pb-Pb collisions is not significant within the current systematic uncertainties, a careful experimental study on the correlated fraction of the systematic uncertainty across different centralities might allow further insights.

Also the deviation of the K$^{0*}$ yield from the thermal model expectation is commonly attributed to the influence of the hadronic phase, but this time to elastic scattering processes of the decay daughters instead of inelastic collisions. The magnitude of the effect is expected to increase with the cross-section of the decay daughters and to decrease for larger lifetimes of the resonance. This picture is now further substantiated by a new ALICE measurement of the $\rho/\pi$ ratio which shows a suppression pattern that follows the ordering expected from the lifetimes~\cite{BelliniProceedings}. 
On the other hand, the results on light (anti-)nuclei production could indicate a very short lifetime of the hadronic phase: despite their low binding energy, which is much smaller than the chemical freeze-out temperature of about 156 MeV, their production yields are in agreement with thermal model expectations. In addition, the shapes of the $p_{\rm T}$-spectra and the elliptic flow pattern are in agreement with a hydrodynamic expansion described by the blast-wave model~\cite{nuclei,Lea:2016jus} and thus show no difference with respect to non-composite particles. Either, this indicates a production at the phase boundary without subsequent interaction, or a formation by final state coalescence (which takes into account the complex space-time dynamics), or an isentropic expansion which forces the system to keep the entropy-per-baryon constant  after chemical freeze-out~\cite{Kapusta:1980zz,Andronic:2010qu}. Simple coalescence models assuming a point-like source without radial expansion fail to describe the data.

\section{Strangeness production at low energies}
In the SPS energy domain, the NA-61/SHINE experiment is on the verge of reaching its originally anticipated goals. The measurement of excitation functions in pp collisions is nearly completed and while data on Ar+Sc collisions is analysed, a beam time for Xe+La is scheduled to complete the system size scan. The data shows that the horn-like structure in the K$^{+}/\pi^{+}$-ratio around $\sqrt{s_{NN}}~\approx~6-10$~GeV (often discussed in the context of thermal models~\cite{Andronic:2008gu} or as a potential signature for the onset of deconfinement~\cite{Alt:2007aa}) is a genuine heavy-ion effect as the pp data follows a flat behaviour. The experiment delivers important input for two external communities. Firstly, with data from proton beams on a carbon target the differential cross-sections for pion production are measured for the precise calculation of the flux of decay neutrinos. Secondly,  data from a pion beam on a carbon target improves the modelling of cosmic ray air showers.


At even lower centre-of-mass energies, the HADES experiment has completed a first comprehensive set of measurements on strange particle production in Au-Au collisions at $\sqrt{s_{NN}} = 2.41$~GeV. Feed-down contributions from $\phi$ decays explain the different slopes of the K$^{+}$ and the K$^{-}$ spectra which were previously attributed to a sequential freeze-out. At such low energies, all strange particles are produced below the nucleon-nucleon threshold leading to a strong sensitivity to medium effects and multi-particle collisions. The data thus helps to constrain better the poorly known kaon-nucleon potential by comparisons with transport codes as shown in Fig.~\ref{fig.:HadesTransport}. The preliminary statistical-thermal fit shows surprisingly that the larger Au+Au system does not follow the trend of the chemical freeze-out curve in contrast to the smaller Ar+KCl system. A future measurement of the $\Xi/\pi$-ratio in the Au+Au system is eagerly awaited in order to see if it also drastically deviates from the thermal model baseline as in the Ar+KCl system~\cite{Agakishiev:2015bwu}.

\begin{figure}
\begin{center}
	\includegraphics[width=0.93\linewidth]{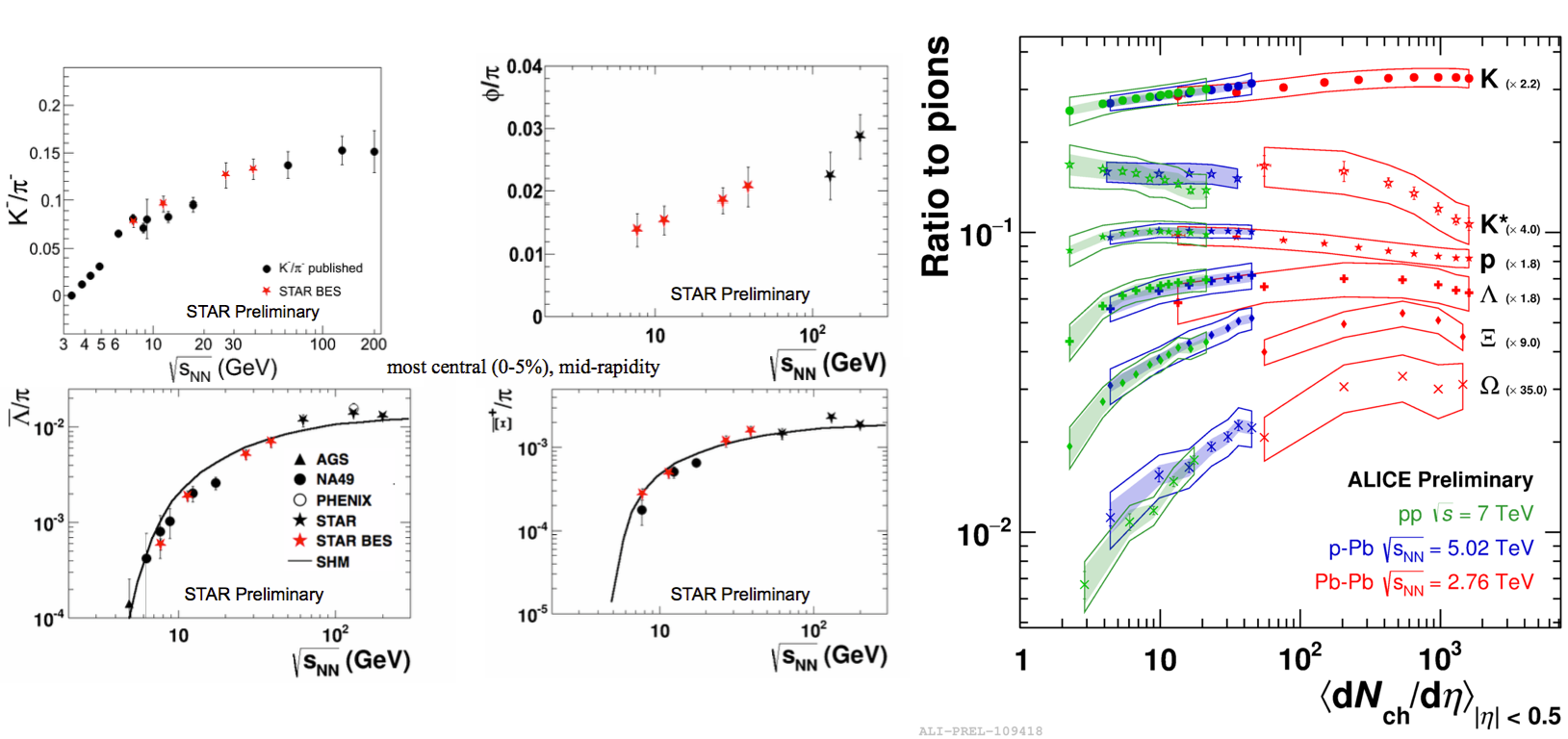}
\end{center}
\caption{\label{fig.:BeamEnergyScanSystemSizeScan} Yields of strange particles as a function of beam energy (left) and as a function of system size at LHC energies (right). Figure taken from~\cite{ShiProceedings} (left) and from~\cite{BelliniProceedings}~(right).}
\end{figure}

\section{Strangeness production in small systems}

Results from the RHIC beam energy scan together with older experimental data nicely demonstrate a smooth evolution of strangeness production (or the particle chemistry in general) across different centre-of-mass energies. One of the most interesting results presented at the conference is that such a smooth evolution of strangeness production is also observed across different collision systems at LHC energies if the multiplicity dependence within the small collision systems is taken into account as shown in Fig.~\ref{fig.:BeamEnergyScanSystemSizeScan}. 
In contrast to p-Pb or Pb--Pb data, the availability of the pp measurements now also allows to compare to a complete set of QCD-inspired event generators. While the DIPSY and EPOS models describe qualitatively the enhanced production of strangeness with increasing event multiplicity, they will need at least further tuning for a correct quantitative description~\cite{BelliniProceedings}. The evolution of the spectral shapes with event multiplicity follows a pattern which is consistent with a common radial expansion. A simultaneous blast-wave fit to the spectra of $\pi$/K/p (ALICE) or $\Lambda$ and K$^{0}_{S}$ (CMS) shows a good description of the spectra in the relevant $p_{\rm T}$-range. The trend of the freeze-out parameters $T_{kin}$ (kinetic freeze-out temperature) and $\langle \beta_T \rangle$ (average radial expansion velocity) is very similar for pp, p--Pb, and Pb--Pb collisions. However, $\langle \beta_T \rangle$ is larger for pp and p--Pb than Pb-Pb collisions at comparable multiplicities~\cite{BelliniProceedings,Khachatryan:2016yru}. This could indicate a stronger radial flow due to larger pressure gradients, if a hydrodynamical description applies to such small systems~\cite{Shuryak:2013ke}.

\section{Summary and roadmap for the future}

A large variety of high quality data on strangeness production from a pp, pA, and AA collisions at various centre-of-mass energies has been presented at the SQM 2016 conference in Berkeley. In the near future, experimental research should address the following five open problems:

\begin{enumerate}

 \item  At the highest RHIC and LHC (run 2) energies, the centrality dependence of yield ratios such as p/$\pi$ and $\Xi$/$\pi$ needs to be measured with good precision including the correlated part of the systematic uncertainty across different centrality bins. This will allow to address the question of potentially different chemical freeze-out temperatures and the influence of inelastic scatterings in the hadronic phase.

 \item Measurements of the event-by-event fluctuations of the conserved quantities in QCD over the entire beam energy range will allow to investigate further a potential flavour dependence of chemical freeze-out and effects related to critical chiral dynamics.

 \item With the silicon vertex detectors operational at RHIC, feed-down corrections for the light flavour hadron yields can be done more precisely. The results should be updated and the thermal-statistical fit should be repeated which might resolve the current discrepancy between the chemical freeze-out temperature at top RHIC energies being slightly higher than the critical temperature obtained in Lattice QCD.
 
 \item The set of strange hadron measurements in Au+Au collisions at SIS-18 energies needs to be completed by the $\Xi$/$\pi$-ratio which showed a surprising deviation in Ar+KCl with respect to the thermal model~\cite{Agakishiev:2015bwu}.

 \item The multiplicity dependence of particle yield ratios in pp collisions at LHC energies should be repeated at 13~TeV centre-of-mass energy in order to disentangle the evolution of particle chemistry with $\sqrt{s}$ from the evolution with event multiplicity.
 
\end{enumerate}

\noindent Even after 25 years of research, the investigation of light flavour hadron production in general and of strange hadrons in particular, remains a highly interesting topic of modern research.

\section*{Acknowledgements}

The author would like to thank A. Andronic, F. Bellini, R. Bellwied, P. Braun-Munzinger, D. Chinellato,  B. Doenigus, R. Holzmann, F. Karsch, H. Oeschler, C. Ratti, K. Safarik, T. Scheib, M. Schmelling, J. Schukraft, S. Shi, J. Steinheimer, H. Stroebele, Y. Zhou.

\section*{References}
\bibliographystyle{iopart-num}
\bibliography{SqmKalweit}

\end{document}